\documentclass[11pt,a4paper]{article}
\usepackage{epsfig}
\usepackage{amssymb}
\usepackage{graphicx}
\usepackage{color}
\usepackage{subfigure}

\usepackage{latexsym}

\begin{document}

\baselineskip 6mm
\renewcommand{\thefootnote}{\fnsymbol{footnote}}

\newcommand{\nc}{\newcommand}
\newcommand{\rnc}{\renewcommand}

%%%%%%%%%%%%%%%%%%%%%% Equation Numbering %%%%%%%%%%%%%%%%%%%%%%%
%\makeatletter \rnc{\theequation}{\thesection.\arabic{equation}}
%\@addtoreset{equation}{section} \makeatother

%%%%%%%%%%%%%%%%%%%%%%%%%%%%%%%%%%%%%%%%%%%%%%%%%%%%%%%%%%%%%%%%%
%                                                               %
%                NEW COMMANDS AND MACROS                        %
%                                                               %
%%%%%%%%%%%%%%%%%%%%%%%%%%%%%%%%%%%%%%%%%%%%%%%%%%%%%%%%%%%%%%%%%

\newcommand{\tcb}{\textcolor{blue}}
\newcommand{\tcr}{\textcolor{red}}
\newcommand{\tcg}{\textcolor{green}}

%%%%% Simplify some frequently used LaTeX commands %%%%%

\def\beq{\begin{equation}}
\def\eeq{\end{equation}}
\def\ba{\begin{array}}
\def\ea{\end{array}}
\def\bea{\begin{eqnarray}}
\def\eea{\end{eqnarray}}
\def\nn{\nonumber}

%%%%%%%%%%%%%%%%%%%%%%%%%%%%%%%%%%%%%
%  Journal 
%%%%%%%%%%%%%%%%%%%%%%%%%%%%%%%%%%%%

\def\CMP{Commun. Math. Phys.~}
\def\JHEP{JHEP~}
\def\Pre{Preprint}
\def\PRL{Phys. Rev. Lett.~}
\def\PR {Phys. Rev.~}
\def\CQG {Class. Quant. Grav.~}
\def\PL {Phys. Lett.~}
\def\NP {Nucl. Phys.~}

%%%%%%%%%%%%%%%%%%%%%%%%%%%%%%%%%%%%%%%%%%%%%%%%%%
%   Boldface Letters
%%%%%%%%%%%%%%%%%%%%%%%%%%%%%%%%%%%%%%%%%%%%%%%%%%
\def\G{\Gamma}

\def\S{{\bf S}}
\def\C{{\bf C}}
\def\Z{{\bf Z}}
\def\R{{\bf R}}
\def\N{{\bf N}}
\def\M{{\bf M}}
\def\P{{\bf P}}
\def\bm{{\bf m}}
\def\bn{{\bf n}}

\def\CA{{\cal A}}
\def\CB{{\cal B}}
\def\CC{{\cal C}}
\def\CD{{\cal D}}
\def\CE{{\cal E}}
\def\CF{{\cal F}}
\def\CM{{\cal M}}
\def\CG{{\cal G}}
\def\CI{{\cal I}}
\def\CJ{{\cal J}}
\def\CL{{\cal L}}
\def\CK{{\cal K}}
\def\CN{{\cal N}}
\def\CO{{\cal O}}
\def\CP{{\cal P}}
\def\CQ{{\cal Q}}
\def\CR{{\cal R}}
\def\CS{{\cal S}}
\def\CT{{\cal T}}
\def\CV{{\cal V}}
\def\CW{{\cal W}}
\def\CX{{\cal X}}
\def\CY{{\cal Y}}
\def\We{{W_{\mbox{eff}}}}

%%%%%%%%%%%%%%%%%%%%%%%%%%%%%%%%%%%%%%%%%%%%%%%%%%
%   Mathematical Symbols  AA
%%%%%%%%%%%%%%%%%%%%%%%%%%%%%%%%%%%%%%%%%%%%%%%%%%

\newcommand{\p}{\partial}
\newcommand{\bp}{\bar{\partial}}

\newcommand{\half}{\frac{1}{2}}

\newcommand{\bfalpha}{{\mbox{\boldmath $\alpha$}}}
\newcommand{\bfbeta}{{\mbox{\boldmath $\beta$}}}
\newcommand{\bfgamma}{{\mbox{\boldmath $\gamma$}}}
\newcommand{\bfmu}{{\mbox{\boldmath $\mu$}}}
\newcommand{\bfpi}{{\mbox{\boldmath $\pi$}}}
\newcommand{\bfvarpi}{{\mbox{\boldmath $\varpi$}}}
\newcommand{\bftau}{{\mbox{\boldmath $\tau$}}}
\newcommand{\bfeta}{{\mbox{\boldmath $\eta$}}}
\newcommand{\bfxi}{{\mbox{\boldmath $\xi$}}}
\newcommand{\bfkappa}{{\mbox{\boldmath $\kappa$}}}
\newcommand{\bfepsilon}{{\mbox{\boldmath $\epsilon$}}}
\newcommand{\bfTheta}{{\mbox{\boldmath $\Theta$}}}

\newcommand{\bz}{{\bar{z}}}

\newcommand{\dalpha}{\dot{\alpha}}
\newcommand{\dbeta}{\dot{\beta}}
\newcommand{\blambda}{\bar{\lambda}}
\newcommand{\btheta}{{\bar{\theta}}}
\newcommand{\bsigma}{{{\bar{\sigma}}}}
\newcommand{\bepsilon}{{\bar{\epsilon}}}
\newcommand{\bpsi}{{\bar{\psi}}}

%%%%%  Temporary notation %%%%

\def\ct{\cite}
\def\la{\label}
\def\eq#1{(\ref{#1})}

%%% Greek letters %%%

\def\a{\alpha}
\def\b{\beta}
\def\g{\gamma}
\def\G{\Gamma}
\def\d{\delta}
\def\D{\Delta}
\def\ep{\epsilon}
\def\e{\eta}
\def\ph{\phi}
\def\Ph{\Phi}
\def\ps{\psi}
\def\Ps{\Psi}
\def\k{\kappa}
\def\l{\lambda}
\def\L{\Lambda}
\def\m{\mu}
\def\n{\nu}
\def\th{\theta}
\def\Th{\Theta}
\def\r{\rho}
\def\s{\sigma}
\def\S{\Sigma}
\def\ta{\tau}
\def\o{\omega}
\def\O{\Omega}
\def\pr{\prime}

%%%%% Mathematical Symbols

\def\half{\frac{1}{2}}

\def\goto{\rightarrow}

\def\na{\nabla}
\def\grad{\nabla}
\def\curl{\nabla\times}
\def\div{\nabla\cdot}
\def\pa{\partial}

\def\bra{\left\langle}
\def\ket{\right\rangle}
\def\lb{\left[}
\def\lc{\left\{}
\def\ls{\left(}
\def\lp{\left.}
\def\rp{\right.}
\def\rb{\right]}
\def\rc{\right\}}
\def\rs{\right)}
\def\cl{\mathcal{l}}

\def\vac#1{\mid #1 \rangle}

%%%%  Special symbol
\def\td#1{\tilde{#1}}
\def\check{ \maltese {\bf Check!}}

%%%%% Roman pont in math

\def\Tr{{\rm Tr}\,}
\def\det{{\rm det}\,}

%%%%% Special format

\def\bc#1{\nnindent {\bf $\bullet$ #1} \\ }
\def\ch {$<Check!>$ }
\def\ss {\vspace{1.5cm}}

\begin{titlepage}

%---------------- preprint number ---------------
\hfill\parbox{5cm} { }

\hskip1cm

% \hskip12cm{CQUeST--2013-0590}

\vspace{10mm}

\begin{center}
%------------------------ title ------------------------
{\Large \bf  Generalized ADT charges  and asymptotic symmetry algebra}

%---------------- authors and addresses ----------------
\vskip 1. cm
  {
 % Wontae Kim$^{ab}$\footnote{e-mail : wtkim@sogang.ac.kr},  
  Shailesh Kulkarni$^a$\footnote{e-mail : shailesh@physics.unipune.ac.in}
  %Sang-Heon Yi$^{a}$\footnote{e-mail : shyi@sogang.ac.kr} 
  }

\vskip 0.5cm

{\it $^a\,$Department of Physics, Savitribai Phule Pune University, Ganeshkhind, Pune, 411007, India }\\
%{\it $^b\,$Department of Physics,, Sogang University, Sinsu-dong, Mapo-gu, Seoul 121-742, Korea}\\

\end{center}

\thispagestyle{empty}

\vskip1.5cm

%----------------------- abstract ----------------------

\centerline{\bf ABSTRACT} \vskip 4mm

\vspace{1cm}
\noindent Using the expressions for generalized ADT  current and potential in a self consistent manner,  we derive the  asymptotic symmetry algebra on AdS$_3$ and  the near horizon extremal BTZ  spacetimes.  The structure of symmetry algebra among the  conserved charges for asymptotic killing vectors matches exactly with the known results thus establishing the algebraic equivalence between the well known existing formalisms for obtaining the conserved charges  and the generalized ADT charges.
\vspace{2cm}
%PACS numbers:

%\today

\end{titlepage}

\renewcommand{\thefootnote}{\arabic{footnote}}
\setcounter{footnote}{0}

%\tableofcontents
%%%%%%%%%%%%%%%%%%%%%%%%%%%%%%
%                                                                            %
%   Sec.  Introduction                                                       %
%                                                                            %
%%%%%%%%%%%%%%%%%%%%%%%%%%%%%%

%%%%%%%%%%%%%%%%%%%%%%%%%%%%%%%%%%%%%%%%%%%%%%%%%%%%%%
\section{Introduction}
%%%%%%%%%%%%%%%%%%%%%%%%%%%%%%%%%%%%%%%%%%%%%%%%%%%%%
Conserved charges in general relativity  and asymptotic symmetries has recently witnessed a plethora of activities 
capturing several interesting aspects of AdS/CFT correspondence \cite{Maldacena:1997re}. The programme initiated with a seminal work of Brown and Henneaux \cite{Brown:1986nw} has now become a corner stone  in establishing a concrete relationship between   symmetry generators of dual conformal field theory (CFT) and asymptotic symmetries of corresponding bulk anti-de-Sitter (AdS) geometry. The formalism relies on the fact that the algebra among the conserved charges for diffeomorphisms satisfying certain asymptotic boundary conditions  is isomorphic to two copies of Virasoro algebra.   Indeed, by extracting the central charge of the Virasoro algebra and using Cardy formula, the  entropy of Banados -Teitelboim-Zanelli (BTZ) black hole was obtained by Strominger \cite{Strominger:1997eq}.  
%The quasi-local conserved charges  in general relativity has always remained an important tool  in understanding of quantum aspects of black holes.  In the context of  AdS/CFT correspondence }  
Since then, there have been much interests in the extension of this work to geometries which are not asymptotically AdS. One such direction has been the study on the asymptotic symmetry algebra in the context of
the Kerr/CFT correspondence \cite{Guica:2008mu, Compere:2012jk}. 

Important step in all the above developments is to identify the conserved charges which form a sub-algebra corresponding to the isometry group of the given spacetime. However,  in theory of gravity with diffeomorphism symmetry, it is not so straightforward to define conserved charges in an unambiguous manner. 
There have been several attempts to define conserved charges in gravity.  In the well known ADM approach developed in \cite{Arnowitt:1962hi}, conserved charges are obtained by introducing on-shell vanishing currents by linearizing the Einstein equations on a given asymptotically flat background. After removing certain ambiguity coming from the potential, one can define conserved quantities through the integral of the on-shell potential corresponding to asymptotic killing vectors.  Extension of  the above method to non-trivial asymptotic boundaries like AdS have been given by Abbott, Deser and Tekin (ADT) \cite{ Abbott:1981ff}.  There are also other approaches to derive quasi-local conserved charges without resorting to the linearization of dynamical fields such as Brown and York method \cite{Brown:1992br} and covariant phase space formalism \cite{ Wald:1993nt, Myers:1993nt, Iyer:1994ys, waldzopus,Papadimitriou:2005ii}. 
General theory of conserved charges based on the cohomology principles  was developed by Barnich-Brandt and Compere (BBC) \cite{Barnich:2001jy, Barnichcomp} (for review see \cite{comp1,Compere:2018aar}). For exact Killing symmetries,  the asymptotic symmetry generators computed within this formalism was shown to be in agreement with those obtained via covariant phase space formalism. 
  
    An alternative approach to understand quasi-local conserved charges has been presented recently \cite{Kim:2013zha}. A key result of our work was the establishment of one-to-one mapping between off-shell linearized Noether potential and ADT potential. Furthermore, by integrating the linearized Noether potential along a path in phase space, we were able to propose a definition of quasi-local conserved charges which are completely consistent with ADT charges.  The method  works well for any covariant theory of gravity including the higher curvature theories and also the gravitational Chern-Simons term \cite{Kim:2013cor}. Extension of this approach  including the  asymptotic Killing vectors has been provided in \cite{ Hyun:2014kfa, Hyun:2014sha, Bhattacharya:2018tse}.
The asymptotic Killing vectors or diffeomorphisms  play a vital role in determining the structure of asymptotic symmetry algebra. Fixing the asymptotic behaviour of fields in gravity is not straightforward as the choice of appropriate fall-off conditions at spatial infinity is not unique. For instance, 
the conventional Brown-Henneaux boundary condition given in \cite{Brown:1986nw, Strominger:1997eq} results into two copies of Virasoro algebra whereas the one used in \cite{Guica:2008mu} leads to chiral Virosoro algebra. Other choices for the boundary conditions  are also possible leading to interesting physics \cite{ Compere:2013bya, Avery:2013dja, Grumiller:2016pqb}. 

Although the application of off-shell ADT method for several interesting geometries is present in the literature \cite{all} (for extensive review see \cite{Adami:2017phg}), its role is limited for exact Killing symmetries.  Thus, it is not apparent as how to perform asymptotic symmetry analysis using off-shell ADT method.  
In order to  strengthen the robustness and wide applicability of off-shell formalism \cite{Kim:2013zha}, it is  natural to look for the construction of symmetry algebras  among the off-shell ADT charges corresponding to the asymptotic diffeomorphisms satisfying  various  boundary conditions  as mentioned in the last paragraph. In this paper, we obtain the asymptotic symmetry algebra entirely within off-shell ADT formalism. To achieve this, we  first derive the  off-shell expressions for quasi-local ADT charges for  the general  asymptotic AdS$_3$ spacetime  which follows from the Brown-Henneaux  boundary conditions \cite{Brown:1986nw}. Then, we give detailed analysis of symmetry algebra satisfied by the generalized off-shell ADT charges. As an another illustration, we apply our method to near horizon extremal BTZ black hole. This geometry provides a naive application of Kerr/CFT correspondence capturing all its essential features. The suitable fall-off conditions are the one given in \cite{Guica:2008mu}. For both the cases, we find that the asymptotic symmetry algebra among the generalized off-shell ADT charges is in complete agreement with  the corresponding results  obtained by using the method given in \cite{Barnich:2001jy, Barnichcomp}.

%%%%%%%%%%%%%%%%%%%%%%%%%%%%%%%%%%%%%%%%%%%%%%%%%%
\section{Generalized ADT charges}\label{GADT}
%%%%%%%%%%%%%%%%%%%%%%%%%%%%%%%%%%%%%%%%%%%%%%%%%%
In this section, we  summarize the method for obtaining off-shell ADT currents and charges given in \cite{Kim:2013zha} and its generalization \cite{Hyun:2014kfa}.   

A generic variation of action for any covariant theory of gravity $S[R, R^2, \cdots]$ is given by
\beq 
\delta S = \frac{1}{16\pi G}\int d^{D}x~    \delta (\sqrt{-g} L ) =    \frac{1}{16\pi G}\int d^{D}x  \Big[  \sqrt{-g} G_{\mu\nu}\delta g^{\mu\nu}+\p_{\mu}\Theta^{\mu}(\delta g) \Big]\,, \label{genVar}
\eeq
where $G^{\mu\nu}=0$ is  the  equations of motion (EOM) for the metric and $\Theta$ is the surface term given by

\beq
\Theta^{\mu}(\delta g) = 2\sqrt{-g}[P^{\mu(\nu\rho)\sigma}\nabla_{\sigma}\delta g_{\nu\rho} - \delta g_{\nu\rho}\nabla_{\sigma}P^{\mu(\nu\rho)\sigma}]\, ,\label{Theta}
\eeq
where the $P-$ tensor is defined as 
\beq
P^{\mu\nu\rho\sigma} = \frac{\p \mathcal{L}}{\p R^{\mu\nu\rho\sigma}} \, .
\eeq
Equating the above variation with the corresponding variation caused by any arbitrary diffeomorphism $\xi$, we identify the off-shell conserved Noether current

\beq J^{\mu}_{\xi} \equiv \p_{\nu}J^{\mu\nu}_{\xi}= 2\sqrt{-g}\, G^{\mu\nu}\xi_{\nu} +  \sqrt{-g}\, \xi^{\mu} L    - \Theta^{\mu}(\xi) \,. \label{Noeth} \eeq

After using Eq.~(\ref{Theta}) in the  last expression  the Noether potential $J^{\mu\nu}_{\xi}$ can be written as

\beq J^{\mu\nu}_{\xi} = \sqrt{-g} \, 2 [P^{\mu\nu\rho\sigma}\nabla_{\rho}\xi_{\sigma} -2 \xi_{\sigma} \nabla_{\rho} P^{\mu\nu\rho\sigma}] \,. \label{Noethpot} \eeq
On the other hand, for  exact Killing symmetries $\zeta$, one can introduce linearized off-shell ADT current $\bar{J}^{\mu}_{ADT}$ and potentials $\bar{J}^{\mu\nu}_{ADT}$ as 

\bea 
     \bar{J}^{\mu}_{ADT}   \equiv \nabla _{\nu}\bar{J}^{\mu\nu}_{ADT} = \delta G^{\mu\nu}\zeta_{\nu} + \frac{1}{2}g^{\alpha\beta}\delta g_{\alpha\beta}\, G^{\mu\nu}\zeta_{\nu}  + G^{\mu\nu}\delta g_{\nu\rho}\, \zeta^{\rho}- \frac{1}{2}\zeta^{\mu}G^{\alpha\beta}\delta g_{\alpha\beta} \, . \label{ADT} 
\eea
Using Bianchi identity and the Killing property of $\zeta$, we can immediately verify the off-shell conservation of $\bar{J}^{\mu}_{ADT}$. It is important to note that the expression for the off-shell ADT current given above is strictly valid for the exact Killing vector $\zeta$ whereas the off-shell Noether potential Eq.~(\ref{Noethpot}) holds true for any diffeomorphism $\xi$. 

The relation between the off-shell ADT potential given above and the off-shell Noether potential follows immediately by considering the variation $\delta g_{\mu\nu}$ around some given background admitting the killing vector $\zeta$. Under this variation, the change in the off-shell Noether current Eq.~(\ref{Noeth}) can be written in terms of the off-shell  Noether potential as

\beq \p_{\nu}(\delta J^{\mu\nu}_{\zeta})= \, 2\delta(\sqrt{-g} G^{\mu\nu}\zeta_{\nu}) + \zeta^{\mu}\delta
(\sqrt{-g}L) - \delta \Theta^{\mu}(g,\delta g) \,. \label{pertnoeth} \eeq
Writing Eq.~(\ref{ADT}) as 
\beq \p_{\nu}(\sqrt{-g}\bar{J}^{\mu\nu}_{ADT})= \, \delta(\sqrt{-g} G^{\mu\nu}\zeta_{\nu}) 
-\frac{1}{2}\sqrt{-g}\zeta^{\mu}G^{\rho\sigma}\delta g_{\rho\sigma}
 \,. \label{ADTpot} \eeq
and invoking the expression for generic variation of $\sqrt{-g}L$ we get

\beq \bar{J}^{\mu\nu}_{ADT}[g,\delta g, \zeta]= \,
\frac{1}{2\sqrt{-g}}\Big[\delta J^{\mu\nu}_{\zeta} - \zeta^{\mu}\Theta^{\nu} + \zeta^{\nu}\Theta^{\mu} \Big]
 \label{ADTkilling}\, . \eeq
Therefore, the infinitesimal change in the conserved ADT charge for a killing vector $\zeta$ is given by
\beq  \delta \bar{Q}_{ADT}(\zeta) = \frac{1}{8\pi G} \int_{\Sigma} d\Sigma_{\mu\nu} \sqrt{-g}\, \bar{J}^{\mu\nu}_{ADT}[g,\delta g, \zeta]\,. \label{deltaADTcharge}
 \eeq
The total  quasi-local conserved charge can be determined by  integrating the above expression along one parameter path in the phase space as

\beq  \bar{Q}_{ADT}(\zeta) = \frac{1}{8\pi G}\int^{1}_{0}d\alpha \int_{\Sigma} d\Sigma_{\mu\nu} \sqrt{-g}\, \bar{J}^{\mu\nu}_{ADT}(g,\delta g, \zeta)\,. \label{ADTcharge}
 \eeq
where $\Sigma$ is a boundary with co-dimension two. 

  We may note that the result for off-shell quasi-local ADT charges given by Eq.~(\ref {ADTcharge}) is limited to exact Killing vectors. However, at the end we are interested in the diffeomorphisms which preserves certain asymptotic boundary conditions. 
 In order to find out the expression for generalized ADT charges corresponding to any asymptotic Killing  vector $\xi$, we consider the right hand side of Eq.~(\ref{ADTkilling}) augmented with $2-$rank antisymmetric tensor $T^{\mu\nu}[g,\delta g, \xi]$ and write the generalized ADT potential as
\beq  {J}^{\mu\nu}_{ADT}[g,\delta g, \xi]= \,
\frac{1}{2\sqrt{-g}}\Big[\delta J^{\mu\nu}_{\xi} - \zeta^{\mu}\Theta^{\nu} + \zeta^{\nu}\Theta^{\mu} \Big] + \frac{1}{2}T^{\mu\nu}[g,\delta g, \xi]. 
 \label{GADT} \eeq

The form for second rank antisymmetric tensor  $T^{\mu\nu}$ appearing in the last equation is restricted by two conditions. First, for exact Killing vector $J^{\mu\nu}_{ADT}$ must match with $\bar{J}^{\mu\nu}_{ADT}$. And secondly, the current built from the above expression is conserved at off-shell level. 
Generic form for $T^{\mu\nu}$ satisfying these two criterion has been derived in \cite{Hyun:2014kfa}
\beq
T^{\mu\nu}[g, \delta g, \xi] = -\frac{1}{2}\Big(3P^{\mu\nu\rho\sigma}g^{\alpha\beta} + 4g^{\sigma[\mu}P^{\nu](\rho\alpha)\beta}\Big)\Big(\nabla_{(\rho}\xi_{\alpha)}\delta g_{\sigma\beta}-  \nabla_{(\sigma}\xi_{\beta)}\delta g_{\alpha\rho}\Big) \label{Ttensor}
\eeq
Therefore, the total conserved charge for arbitrary diffeomorphism $\xi$ is given by
\beq  Q_{ADT}(\xi) =  \frac{1}{8\pi G}\int^{1}_{0}d\alpha \, \delta Q[g,\delta g, \xi] = \frac{1}{8\pi G}\int^{1}_{0}d\alpha \int_{\Sigma} d\Sigma_{\mu\nu} \sqrt{-g}\, J^{\mu\nu}_{ADT}(g,\delta g, \xi)\,. \label{ADTcharge2}
 \eeq
We shall use this generalized ADT charge valid for any asymptotic Killing  diffeomorphism  in order to derive the asymptotic symmetry algebra. 
\section{Asymptotic symmetry algebra}
We now  turn our attention towards  computation of asymptotic symmetry algebra satisfied by the generalized ADT charges Eq.~(\ref{ADTcharge2}). 
We illustrate our construction for the diffeomorphisms  which preserves two well known boundary conditions given in \cite{ Brown:1986nw} and \cite{Guica:2008mu}. For definiteness, we  concentrate on $3-$d Einstein gravity described by the action
\beq
 S [g_{\mu\nu}]  = \frac{1}{16\pi G}\int d^{3}x \sqrt{-g}\Big[R + \frac{2}{\ell^2} \Big] \,. 
\eeq
First we note that for the above theory the $P-$ tensor can be written as
\beq
P^{\mu\nu\rho\sigma} = \frac{1}{2}(g^{\mu\rho}g^{\nu\sigma} - g^{\mu\sigma}g^{\nu\rho})\, . \label{PtensorEH}
\eeq
Hence, the expression for off-shell Noether potential $J^{\mu\nu}_{\xi}$ becomes
\bea J^{\mu\nu}_{\xi} &=& \sqrt{-g} \, [\nabla^{\mu}\xi^{\nu} - \nabla^{\mu}\xi^{\nu}] \,. \label{NoethpotEinst}
 \eea
Taking the $\delta$ variation and denoting $\delta g_{\mu\nu} =h_{\mu\nu}$, we get
\bea
\delta J^{\mu\nu}_{\xi} &=& \sqrt{-g}
\Big\{h^{\mu\beta}\nabla_{\beta}\xi^{\nu}  - h^{\nu\beta} \nabla_{\beta}\xi^{\mu}
+ g^{\beta[\mu}g^{\nu]\rho}(\nabla_{\beta}h_{\rho\alpha} + \nabla_{\alpha}h_{\beta\rho} - \nabla_{\rho}h_{\beta\alpha})\xi^{\alpha} \nn \\
&& + h \nabla^{[\mu}\xi^{\nu]} \Big\} \nn \\
&=& 2\sqrt{-g}\Big[\xi_{\alpha}\nabla^{[\mu}h^{\nu]\alpha} - h^{\alpha[\mu}\nabla_{\alpha}\xi^{\nu]} + \frac{1}{2}h\nabla^{[\mu}\xi^{\nu]}\Big]\, .
\eea
Also, the surface term in  Eq.~(\ref{Theta}) simplifies to 
\beq
\Theta^{\mu}[g, \delta g] = \sqrt{-g}[\nabla^{\mu}(g_{\rho\sigma}\delta g^{\rho\sigma}) - \nabla_{\nu}\delta g^{\mu\nu}] = \sqrt{-g}[ \nabla_{\nu}h^{\mu\nu} - \nabla^{\mu} h]\, .
\eeq
Therefore, the first term in Eq.~(\ref{GADT}) can be written as 
\bea
\delta J^{\mu\nu}_{\xi} - \zeta^{\mu}\Theta^{\nu} + \zeta^{\nu}\Theta^{\mu} &=& 
2\sqrt{-g}\Big[ \xi_{\alpha}\nabla^{[\mu}h^{\nu]\alpha} - h^{\alpha[\mu}\nabla_{\alpha}\xi^{\nu]} + \frac{1}{2}h\nabla^{[\mu}\xi^{\nu]} - \xi^{[\mu}\nabla_{\alpha}h^{\nu]\alpha} \nn\\
&& + \, \xi^{[\mu}\nabla^{\nu]}h\Big]\, .
\eea

 Using the above  equation in Eq.~(\ref{GADT}) and substituting Eq.~(\ref{PtensorEH})  into Eq.~(\ref{Ttensor}) we get the expression for generalized ADT potential for Einstein gravity 
\bea
2\sqrt{-g}J_{ADT}^{\mu\nu}[g,\delta g, \xi] &=& 2\sqrt{-g}\Big\{ \Big[\xi_{\alpha}\nabla^{[\mu}h^{\nu]\alpha} - h^{\alpha[\mu}\nabla_{\alpha}\xi^{\nu]} + \frac{1}{2}h\nabla^{[\mu}\xi^{\nu]} - \xi^{[\mu}\nabla_{\alpha}h^{\nu]\alpha} + \xi^{[\mu}\nabla^{\nu]}h \Big] \nn\\
&& - \frac{1}{2}\Big(h^{\rho[\mu}\nabla_{\rho}\xi^{\nu]} - h^{\rho[\mu}\nabla^{\nu]}\xi_{\rho}\Big) \Big\}\, . \label{GADTeinstein}
\eea
Note that  in this expression for generalized ADT potential the fields $g_{\mu\nu}$ and its perturbations need not have to satisfy the equations of motion. Also, the diffeomorphism $\xi$ is not restricted to exact symmetry. 
%\bea
%\frac{1}{2}\Big[ h \nabla^{[\mu}\xi^{\nu]}  - h^{\alpha[\mu}\nabla_{\alpha}\xi^{\nu]} + h^{\alpha[\mu}\nabla^{\nu]}\xi_{\alpha}\Big] \nn \\
%&& - \xi^{[\mu}\nabla_{\alpha}h^{\nu]\alpha} + \xi_{\alpha}\nabla^{[\mu}h^{\nu]\alpha}
%+ \xi^{[\mu}\nabla^{\nu]}h  \label{GADTeinstein}
%\eea
\subsection{Brown-Henneaux conditions}
Any asymptotically AdS$_3$ solution for Einstein equations near the boundary can be expanded in Feffermann-Graham coordinates \cite{Balasubramanian:1999re} as
\beq
ds^2 = \frac{\ell^2 dr^2}{r^2} + \frac{r^2}{\ell^2} \Big(g_{(0)ab}  + \frac{\ell^2}{r^2} g_{(2)ab}  +  \frac{\ell^4}{r^4} g_{(4)ab} \Big) dx^a dx^b \, ,  \label{feff}
\eeq
with $x^a = x^{\pm} = \frac{t}{\ell} \pm \phi$ and $\phi = \phi+2\pi$.\\
We now look for the diffeomorphisms which change the metric near the asymptotic boundary as 
\bea
g_{rr} &=& \frac{1}{r^2} + \mathcal{O}(\frac{1}{r^4}) \qquad g_{tt} =  -r^2 + \mathcal{O}(1) \qquad g_{tr}= \mathcal{O}(\frac{1}{r^3}) \nn \\
g_{t\phi} &=& \mathcal{O}(1) \ \qquad \qquad  \ g_{r\phi} = \mathcal{O}(\frac{1}{r^3}) \ \qquad g_{\phi\phi} =  r^2 + \mathcal{O}(1)\, . 
\eea
In Fefferman-Graham coordinates above set of conditions simply become
\bea
g_{(0)++} = g_{(0)- -} = 0  \qquad \ g_{(0)+ -} = g_{(0)- +}  = -\frac{1}{2} \label{BHC}\, ,
\eea
with subleading terms unconstrained. 
The diffeomorphisms  $\xi^{\mu}$  which preserves Eq.~(\ref{BHC}) are given by \cite{Strominger:1997eq}
\bea
\xi^{+}(x^{+}) &=& f(x^+)\p_{+} -\frac{1}{2} \p_{+}f(x^{+}) r\p_{r} + \frac{\ell^2}{2r^2}\p_{+}^{2}f(x^{+})\p_{-} \nn \\
\xi^{-}(x^{-})   &=& g(x^-)\p_{-} -\frac{1}{2} \p_{-}g(x^{-}) r\p_{r} + \frac{\ell^2}{2r^2}\p_{-}^{2}g(x^{-})\p_{+}\, ,
\eea
where $f(x^{+})$ and $g(x^{-})$  are arbitrary function of their arguments. Since $\phi$ is $2\pi$ periodic,  we can expand the above asymptotic diffeomorphisms in Fourier modes $e^{i m x^{\pm}}$ and write 
\bea
\xi^{+}_{m}(x^{+}) &=& e^{i m x^{+}}\Big[\p_{+} -\frac{i m r}{2}\p_r - \frac{\ell^2 m^2}{2r^2}\p_{-}\Big] \nn \\
\xi^{-}_{m}(x^{-}) &=& e^{i m x^{-}}\Big[\p_{+} -\frac{i m r}{2}\p_r - \frac{\ell^2 m^2}{2r^2}\p_{+}\Big]\, . \label{difffour}
\eea
It is easy to see that the above diffeomorphisoms satisfy Diff($S^{1})$ algebra
\beq
\{ \xi^{\pm}_{m}(x^{\pm}), \xi^{\pm}_{n}(x^{\pm}) \} = i(n-m)\xi_{m+n}^{\pm} \label{DiffS1}
\eeq
To obtain the general form for $g_{(2)ab}$,  we solve the Einstein equation order by order. This fixes the form for components  of  $g_{(2)ab}$ in terms of two arbitrary functions $T(x^{+})$ and $\bar{T}(x^{-})$  as 
\bea
g_{(2)+ -} &=&  g_{(2)- +}  = 0 \qquad \ g_{(2)+ +} = \ell^2 T(x^{+}) \qquad \  g_{(2) - -} = \ell^2 \bar{T}(x^-)\, ,
\eea
while the subleading terms are  constrained by the leading order ones as  
\beq
 g_{(4)ab} = \frac{1}{4}g_{(2)ac}g_{(0)}^{cd}g_{(2)db}\, .
\eeq
Therefore, the most general form for solution of Einstein equations respecting the Brown-Henneaux boundary conditions is given by
\bea
g_{\mu\nu} = \left({\begin{array}{ccc}
\frac{\ell^2}{r^2}&0&0\\ 
0& \ell^2 T(x^+)&  -\frac{1}{2}\Big[r^2 + \frac{\ell^4}{r^2}T(x^+)\bar{T}(x^-)\Big]\\
0& -\frac{1}{2}\Big[r^2 + \frac{\ell^4}{r^2}T(x^+)\bar{T}(x^-)\Big]& \ell^2 \bar{T}(x^-)\\
\end{array}}
\right) \label{generalmetricBH}
\eea
 By setting $T^{+} + T^{-} = 4GM$ and $T^{+} + T^{-} = \frac{4GJ}{\ell}$,  we recover the well known BTZ (rotating) black hole  metric (in $t,r,\phi$ coordinates), as
\bea 
ds^2 &=& -N^2 dt^2 +\frac{dr^2}{N^2} + r^2 (d\phi + N_{\phi}dt)^2\, \\
 N^2 &=& -8MG + \frac{r^2}{l^2} + \Big(\frac{4GJ}{r}\Big)^2  \ ; \ N_{\phi} = -\frac{4GJ}{r^2} \nn
\eea
 We would like to stress that for off-shell ADT formalism to be self-consistent one must identify $M$ and $J$  with the conserved charges derived  from the off-shell expression Eq.~({\ref{ADTcharge}) corresponding to the exact killing vectors $\zeta = \p_{t}$ and $\xi = \p_{\phi}$, respectively. 
To this end, we now calculate the mass $M$ for the BTZ solution. 
 For the timelike Killing vector $\zeta = \p_{t}$ the Eq.~(\ref{deltaADTcharge}) takes the form
\beq
\delta \bar{Q}_{ADT}(\p_t) = \frac{1}{16\pi G}\int_{\Sigma} \,  d\Sigma_{rt} \, \Big[\delta \bar{J}^{rt} + \zeta^{t} \Theta^{r}\Big]\, .
\eeq
For Einstein gravity the $P-$ tensor become divergenceless. As a result, the  contribution from $\delta J^{\mu\nu}_{\p_t}$  to $\bar{Q}(\p_t)$ vanishes. Therefore, 
 \beq
\bar{Q}_{ADT}(\p_t) = \frac{1}{16\pi G}\int_{0}^{1} \, d\alpha \, \int_{\Sigma} \,  d\Sigma_{rt} \,  \zeta^{t} \Theta^{r}[g, \delta g]\, .
\eeq
 Writing the perturbation around the background as 
\beq
\displaystyle \delta g_{\mu\nu} = \frac{\p g_{\mu\nu}}{\p \alpha}\delta \alpha \ ; \ \delta \alpha =  G \delta M.
\eeq
and evaluating $r$ component of  Eq.~(\ref{Theta})  we get
\beq  \delta \bar{Q}_{ADT}(\p_t) = \frac{\delta \alpha }{G} = \delta M\, . 
 \eeq
The total charge  is obtained by integrating over $\alpha$ from $0$ to $\alpha >0$ (that is varying $M$ from $0$ to any final value), is given by 
\beq
\bar{Q}_{ADT}(\p_t) = M
\eeq
Similar computation can be performed for the rotational Killing vector to obtain the angular momentum of BTZ black hole as
 \beq
\bar{Q}_{ADT}(\p_\phi) = J\, .
\eeq
Thus, we have obtained the desired expressions for mass and angular momentum for BTZ metric within the  off-shell formalism. 

We are now in a position to compute the symmetry algebra among the generalized ADT charges. 
The infinitesimal variation of ADT charges for the diffeomorphisms Eq.~(\ref{difffour}) is given by
\footnote{From now onwards we drop the subscript ADT from the expressions for ADT charges.}
\beq  \delta Q (\xi^{\pm}_{m}) \equiv \delta Q^{\pm}_{m} =\frac{1}{8\pi G} \int_{\Sigma} d\Sigma_{\mu\nu} \sqrt{-g}\, J_{ADT}^{\mu\nu}[g,\delta g, \xi^{\pm}_{m}]\,. \label{GADTchargefour}
 \eeq
% we
%get the expression for generalized ADT potential
%\bea
%J^{\mu\nu}_{ADT}[g,\delta g, \xi^{\pm}_{m}] &=& \Big[\frac{1}{2} h \nabla^{[r}(\xi^{\pm}_{m})^{+]}  
%- (\xi^{\pm}_{m})^{[r}\nabla_{\alpha}h^{+]\alpha} + (\xi^{\pm}_{m})_{\alpha}\nabla^{[r}h^{+]\alpha}
%+ (\xi^{\pm}_{m})^{[r}\nabla^{+]}h\Big] \nn\\
%&&- \frac{1}{2}\Big(h^{\alpha[r}\nabla_{\alpha}(\xi^{\pm}_{m})^{+]} + h^{\alpha[r}\nabla^{+]}(\xi^{\pm}_{m})_{\alpha}\Big) \nn \\
%&&   \label{GADTkillingmodes}
%\eea
Since the boundary $\Sigma$ is a circle, we only need to compute $(r +)$ and $(r -)$ components of  $J^{\mu\nu}[g,\delta g]$. By explicitly calculating these relevant components for the diffeomorphism $\xi^+$, we find 
 \bea
 \delta Q^{+}_{m}  &=& \frac{1}{8\pi G} \int_{0}^{2\pi}\, d\phi \sqrt{-g}\, \Big( \epsilon_{r+\phi}J^{r+}[g,\delta g, \xi^{+}] +  \epsilon_{r-\phi}J^{r-}[g,\delta g, \xi^{+}] \Big)\, \nn\\ 
&=&  \frac{1}{8\pi G} \int_{0}^{2\pi}\, d\phi \, \Big\{ \Big[\frac{1}{2} h \nabla^{[r}(\xi^{+}_{m})^{+]}  
- (\xi^{+}_{m})^{[r}\nabla_{\alpha}h^{+]\alpha} + (\xi^{+}_{m})_{\alpha}\nabla^{[r}h^{+]\alpha}
+ (\xi^{+}_{m})^{[r}\nabla^{+]}h\Big] \nn\\
&&- \frac{1}{2}\Big(h^{\alpha[r}\nabla_{\alpha}(\xi^{+}_{m})^{+]} + h^{\alpha[r}\nabla^{+]}(\xi^{+}_{m})_{\alpha}\Big) \Big\} + \Big\{ + \rightarrow -\Big\} \, .
\label{deltaGADTfinal}
 \eea
In the above expression,  the first term in the curly bracket is coming from $(r+)$ component while the the second curly bracketed term (denoted by $+ \rightarrow -$) is due to $(r-)$  component. This term is same as the first one with $+$ component replaced by $-$.  Also, note that in the square bracketed expression in the first curly bracket originates from the first part on the right hand side of Eq.~(\ref{GADTeinstein}). 
 
 For the metric Eq.~(\ref{generalmetricBH}) perturbations $h_{\mu\nu}$ induced by $\delta T(x^{+})$ and $\delta \bar{T}(x^{-})$ can be written as
\beq
h_{\mu\nu} = \frac{\p g_{\mu\nu}}{\p T}\delta T(x^{+}) + \frac{\p g_{\mu\nu}}{\p \bar{T}}\delta \bar{T}(x^{-})\, .
\eeq
The nonvanishing components for metric perturbations are given by
\bea
h_{++} =\ell^2 \delta T \  \ h_{--} = \ell^2 \delta \bar{T}\ \ h_{+ -} = h_{-+} =-\frac{\ell^4}{2r^2}(\bar{T} \delta T + T \delta\bar{T})\, .
\eea
Substituting these expressions in Eq.~(\ref{deltaGADTfinal}) we obtain
\bea
 \delta Q^{+}_{m}  = \frac{\ell}{8\pi G}\int_{0}^{2\pi} \, d\phi e^{i m x^{+}} \delta T(x^+)\,. \label{deltaGADT+2}
 \eea
Similarly for $\xi^{-}$ we get
 \bea
 \delta Q^{-}_{m} =  \frac{ \ell}{8\pi G}\int_{0}^{2\pi} \, d\phi e^{i m x^{-}} \delta T(x^-)\,. \label{deltaGADT-}
 \eea
It is clear that the above expression is integrable. Setting the functions $T$ and $\bar{T}$ to zero for the background (zero mass solution) we get the expressions for total charge for the respective asymptotic diffeomorphisms $\xi^{+}$ and $\xi^{-}$ as
\bea
Q_{m}^{+} &=& \frac{\ell}{8\pi G} \int_{0}^{2\pi} d\phi \, T(x^{+}) e^{i m x^{+}} \nn \\
Q_{m}^{-} &=&\frac{\ell}{8\pi G} \int_{0}^{2\pi} d\phi \, \bar{T}(x^{-}) e^{i m x^{-}}\, .
\eea
For any two diffeomorphisms the Lie bracket among the corresponding conserved charges is defined as 
\beq
\{Q_{m}^{\pm}, Q_{n}^{\pm} \} = \delta_{\xi^{\pm}_{n}}Q_{m}^{\pm}\label{algebra}
\eeq
In particular, for $\xi^{+}$ we have
\bea
\delta_{\xi^{+}} Q_{m}&=& \frac{\ell}{8\pi G} \int_{0}^{2\pi} d\phi \, [\mathcal{L}_{\xi^{+}_{n}}T(x^{+})] e^{i m x^{+}}\, . \label{delQ+}
\eea
The change in $T(x^+)$ between two nearby solutions $g$ and $g+\delta_{\xi}g$  is given by
\bea
\mathcal{L}_{\xi^{+}_{n}}T(x^{+})  = e^{i n x^{+}}\Big(\p_{+} + 2 i n + \frac{i n^3}{2} \Big)T(x^+)\, . \label{deltaT+}
\eea
Using Eq.~(\ref{delQ+}) together with  Eq.~(\ref{deltaT+}) in Eq.~(\ref{algebra}), we finally get
\bea
\{Q_{m}^{+}, Q_{m}^{+} \} &=& i(n-m) Q_{m+n}^{+}+ \frac{\ell}{8G}m^3 \delta_{m,-n}\, \nn \\
\{Q_{m}^{-}, Q_{m}^{-} \} &=& i(n-m) Q_{m+n}^{-}+ \frac{\ell}{8G}m^3 \delta_{m,-n}\, .
\eea
This is nothing but the two copies of the Virasoro algebra  with the central charge $c$ given by
\beq
c = \frac{3\ell}{2G}\, .
\eeq
Thus, we have shown that the algebra of generalized ADT charges is consistent with that obtained in \cite{Brown:1986nw, Strominger:1997eq}.\\ 
\subsection{Near horizon extremal BTZ }
As an another application of off-shell method, we now construct the symmetry  algebra among the symmetry for  the near horizon geometry of extremal BTZ black hole spacetime. The asymptotic 
symmetry analysis of this geometry is well studied in the literature in the context of Kerr/CFT correspondence
\cite {Guica:2008mu}. The phase space to which the near horizon extremal BTZ geometry belongs is not the same as  the one represented by Eq.~(\ref{generalmetricBH}). Consequently,  the boundary condition and the corresponding asymptotic diffeomorphisms are quite different as compared to the usual Brown-Henneaux conditions discussed in the previous subsection. Therefore, it is important to study the asymptotic symmetry algebra using generalized ADT 
 charges and to see whether it matches with  the known results. 

 The extremal BTZ black hole is described by the line element in Fefferman-Graham coordinates as
\beq
ds^2 = \frac{\ell^2}{r^2}dr^2 - \Big(r dx^{-} - \frac{4 \ell T(x^{+})}{r}dx^{+}\Big)\Big(r dx^{+} - \frac{4\ell \bar{T(x^{-})}}{r}dx^{-}\Big) 
\eeq
where $\ell M = T(x^{+}) + \bar{T}(x^{-})$ and $J= T(x^{+}) - \bar{T}(x^{-})$. 
Near horizon limit of the  extremal BTZ black hole can be obtained by considering the following coordinate transformations 
\beq
t = \frac{\tau\sqrt{J\ell}}{\alpha} \ ; \ r = \ell \sqrt{\alpha\rho} \ ; \ \phi = \tilde{\phi} - \frac{\tau\sqrt{\ell J}}{\ell\alpha}\, ,
\eeq
and taking $\alpha \rightarrow 0$ limit. Thus, in the near horizon limit of the extremal BTZ black hole 
$x^{+} = \tilde \phi$ is well defined but $x^{-}$ diverges. This can be resolved by setting $\bar{T} (x^{-})=0$.  In the new coordinates, the near horizon extremal BTZ black hole metric is described by
\beq
ds^{2} = \frac{\ell^2}{4}\Big[\frac{d\rho^{2}}{\rho^2} - \rho^2 d\tau^{2} +
r_{H}^{2}\Big(d\tilde{\phi} - \frac{\rho d\tau}{r_{H}}\Big)^2 \Big] \ ; \ r_{H} = 4\sqrt{\frac{J}{\ell}} \label{NHEBTZ}
\eeq
This 
Now we impose the asymptotic boundary conditions 
\bea
g_{\tau\tau} &=& \mathcal{O}(\rho^2) \qquad  \ g_{\tau\rho} =  \mathcal{O}(\frac{1}{\rho^2}) \qquad \ g_{\tau\tilde{\phi}} = -\frac{r_{H}\ell^{2} \rho}{4}+ \mathcal{O}(1) \nn \\
g_{\rho\rho} &= & \frac{\ell^2}{4\rho^2}+ \mathcal{O}(\frac{1}{\rho^3}) \qquad  \  g_{\rho\tilde{\phi}} =  \mathcal{O}(\frac{1}{\rho}) \qquad \ g_{\tilde{\phi}\tilde{\phi}} =  \mathcal{O}(1)\, .\label{NHEBTZBC}
\eea
These conditions can be regarded as $3-$ dimensional analogue of the fall-off conditions used in Kerr/CFT correspondence \cite{Guica:2008mu, Compere:2012jk}.  Recently, it has been argued that the geometry Eq.~(\ref{NHEBTZ}) can be thought of as the decoupling limit of phase space of all Banados geometries with vanishing  left (or right) moving modes \cite{Compere:2015knw}. 
Note the the geometry given by Eq.~(\ref{NHEBTZ}) admits $SL(2,\mathbb{R}) \times U(1)$ isometry group and the boundary conditions Eq.~(\ref{NHEBTZBC}) enhances the $U(1)$ part of the full isometry group. The asymptotic symmetry generator that preserves the above boundary conditions is given by 
\beq
\xi_{m} = e^{- i m \tilde \phi}\Big(-\p_{\tilde{\phi}} - i m \rho \p_{\rho}\Big)\, .
\eeq
It is easy to show that the above right moving modes $\xi_{m}$ satisfy Diff($S^1$) algebra Eq.~(\ref{DiffS1}) as before. Substituting $\xi_{m}$ in  Eq.~(\ref{GADTeinstein})  and computing the relevant, ($\tau, r$) components, we find 
\bea
J^{\tau\rho} = \frac{1}{4}\Big[2 i \, m\,  \rho \, h_{\rho\tilde{\phi}} - \Big(\frac{h_{\tau\tau}}{\rho^2} + 2h_{\tilde{\phi}\tilde{\phi}} + 2\rho \partial_{\tilde{\phi}}h_{\rho\tilde{\phi}}\Big) \Big] + \cdots \label{trADT}
\eea
with the dots representing the terms which vanish for large $\rho$.
Taking Eq~(\ref{NHEBTZ}) as a background, we explicitly find the expressions for metric perturbations 
\bea
h_{\tau\tau} &=& \mathcal{L}_{\xi_{m}}g_{\tau\tau} = 0 \ ;  \  h_{\rho\rho} = \mathcal{L}_{\xi_{m}}g_{\rho\rho} = 
-\frac{G\ell^2}{2(1+\rho^2)^2} i m e^{- i m \tilde{\phi}} \nn \\
h_{\rho\phi} &= &\mathcal{L}_{\xi_{m}}g_{\rho\tilde{\phi}} = -\frac{G\ell^2 \rho}{4(1+\rho^2)}  m^2 e^{- i m \tilde{\phi}} \ ; \ 
h_{\tilde{\phi}\tilde{\phi}} = \mathcal{L}_{\xi_{m}}g_{\tilde{\phi}\tilde{\phi}} = \frac{G\ell^2}{2}i \  m e^{-i m \tilde{\phi}}\, .
\eea
Substituting the above expressions in Eq.~(\ref{trADT}) and integrating the above expression over a circle  at $\rho\rightarrow \infty$ \footnote{ Here we have assumed the infinitesimal the integrability  of the surfaces charge $\delta Q_{m}$. } we get the central extension as
\beq
\{Q(\xi_m), Q(\xi_{n})\} - Q({\{\xi_{m},\xi_{n}\}}) = \frac{1}{8\pi G}\int_{0}^{2\pi} \, d\tilde{\phi}\,  K_{\xi_{m}}[\mathcal{L}_{\xi_{n}}g, g] =  -i  \frac{G\ell m}{8}\Big[m^2 + r_{H} \Big] 
\eeq
%\bea
%J^{\tau\rho}[\xi_m] = -i m \frac{G\ell^2}{8}\Big[\frac{2\rho^2}{(1+\rho^2)}m^2 + m\Big] + \mathcal{O}(\frac{1}{\rho})\, .
%\eea
 This gives the central charge  $c =\frac{3\ell}{2G}$  same as  Brown-Henneaux charge discussed earlier. The difference is that in this case the algebra among the charges is chiral, with only one Virasoro symmetry.  Our result  matches with the previously known results \cite{ Compere:2015knw, Azeyanagi:2008dk} in which the charges were obtained by using the method given in \cite{Barnich:2001jy,Barnichcomp}.  
\section{Conclusion}
In this work, we have given an alternative method based on the off-shell ADT formalism to obtain the asymptotic symmetry algebra.  The similarity between the generalized ADT approach initiated in \cite{Kim:2013zha} and other existing methods was mentioned in \cite{Hyun:2014kfa}. In particular, the expressions for quasi-local conserved charges for any asymptotic Killing vectors obtained in off-shell ADT formalism bear a close resemblance with the BBC formalism \cite{comp1}. However, the construction of symmetry algebra within  off-shell ADT formalism has not been performed. We have provided explicit equivalence between the two approaches in a consistent manner by  computing  the asymptotic symmetry algebra among the generalized ADT charges for asymptotic AdS$_3$ and near horizon extremal BTZ geometries. 
In the first case, we have considered the most general asymptotic AdS$_3$ spacetime  dictated  by the fall-off conditions given in \cite{Brown:1986nw}. Working in the Fefferman-Graham coordinates, we have explicitly
computed the expression for generalized ADT surface charges. The algebra among these charges was shown to reproduce two copies of Virasoro algebra. This confirms that the off-shell formalism for quasi-local conserved charges is equivalent to BBC approach \cite{Barnich:2001jy, Barnichcomp}  even at the level of asymptotic symmetry algebra. It is worthwhile to mention that by construction, the generalized ADT charges is  valid for any  arbitrary background which may or may not satisfy the field equations. However, while computing the asymptotic symmetry algebra, one has to take  background metric as well as its perturbations, both satisfying  the equations of motion. The algebraic equivalence between the two methods is mainly because  in the Fefferman-Graham gauge, the difference between the relevant components of the generalized ADT potential and Barnich-Brandt's $2-$ form potential vanishes. Consequently,   the structure of asymptotic symmetry algebra remain unaltered. We would like to point out that the similar simplification occurs in BBC formalism also. In fact, for Einstein gravity one can explicitly show that the difference between the Barnich-Brandt's surface charge and Noether-Wald surface charge vanishes in the  Fefferman-Graham gauge. To further strengthen this equivalence, we have constructed the symmetry algebra for near horizon extremal BTZ spacetime.  In this case too, our findings are in exact agreement with existing  literature \cite{Azeyanagi:2008dk} wherein the asymptotic symmetry algebra was obtained by using BBC approach.

   Here, we have used Einstein gravity as a prototype example to establish the correspondence  between the above mentioned approaches for the computation of quasi-local charges. It is interesting exercise to extend this algebraic correspondence for the gravity theories containing Chern-Simons term. 
Apart from the boundary conditions used in our work there are other interesting conditions giving rise to deeper understanding of asymptotic symmetry analysis \cite{Grumiller:2016pqb, Azeyanagi:2011zj}. It would be also interesting to incorporate these boundary conditions within generalized ADT formalism. 
We would like to address these issues in the near future.  

%\subsection{And subsequent}
%\subsubsection{Sub-sections}
%\paragraph{Up to paragraphs.} We find that having more levels usually
%reduces the clarity of the article. Also, we strongly discourage the
%use of non-numbered sections (e.g.~\texttt{\textbackslash
%  subsubsection*}).  Please also see the use of
%``\texttt{\textbackslash texorpdfstring\{\}\{\}}'' to avoid warnings
%from the hyperref package when you have math in the section titles

%\appendix
%\section{Some title}
%Please always give a title also for appendices.
\section{acknowledgments}
I would like to thank Dileep Jatkar  and Sang-Heon Yi for the fruitful discussions and correspondence. This work is supported by INSPIRE faculty scheme (research grant No.IFA-13 PH-56) by Department of Science and Technology (DST), Govt. of India and UGC-Faculty Recharge Programme  Govt. of India.

%%%%%%%%%%%%%%%%%%%%%%%%%%%%%%%%%%%%%%%%%%%%%

 %%%%%%%%%%%%%%%%%%%%%%%%%%%%%%%%%%%%
\end{document}